\documentclass[twocolumn,showkeys,aps,prl,showpacs]{revtex4-1}
\usepackage{graphicx}
\usepackage[CJKbookmarks,dvipdfm,colorlinks,linkcolor=blue,citecolor=blue]{hyperref}
\usepackage{amsmath, amssymb}
\usepackage{makecell}
\usepackage{multirow}
\usepackage{CJK}
\usepackage{mathrsfs}
\usepackage{bm}
\usepackage{amsmath}
\usepackage{dcolumn}
\usepackage{epstopdf}
\usepackage{dsfont}
\usepackage{amssymb}
\usepackage{tabularx}
\usepackage{array}
\usepackage{float}
\usepackage{color}
\usepackage{epstopdf}
\usepackage{mathrsfs}
\usepackage{extarrows}

\begin{document}

\title{Piezoelectric altermagnetism and spin-valley polarization in Janus monolayer $\mathrm{Cr_2SO}$}

\author{San-Dong Guo$^{1}$, Xiao-Shu Guo$^{1}$, Kai Cheng$^{1}$, Ke Wang$^{1}$  and Yee Sin Ang$^{2}$}
\affiliation{$^1$School of Electronic Engineering, Xi'an University of Posts and Telecommunications, Xi'an 710121, China}
\affiliation{$^2$Science, Mathematics and Technology (SMT), Singapore University of Technology and Design (SUTD), 8 Somapah Road, Singapore 487372, Singapore}

\begin{abstract}
The altermagnetism can achieve spin-split bands in collinear symmetry-compensated antiferromagnets. Here, we predict  altermagnetic order in Janus monolayer $\mathrm{Cr_2SO}$ with eliminated inversion symmetry, which can realize the combination of piezoelectricity and altermagnetism in a two-dimensional  material, namely 2D piezoelectric altermagnetism. It is found that $\mathrm{Cr_2SO}$ is an altermagnetic semiconductor, and the spin-split bands of both valence and conduction bands are near the Fermi level. The $\mathrm{Cr_2SO}$ has large out-of-plane piezoelectricity ($|d_{31}|$$=$0.97 pm/V), which is highly desirable for ultrathin piezoelectric device application.
 Due to spin-valley locking, both spin and valley can be polarized by simply breaking the corresponding crystal
symmetry with uniaxial strain.  Our findings provide a platform
to integrate spin, piezoelectricity and valley in a single material, which is useful for multi-functional device applications.

\end{abstract}
\keywords{Altermagnetism, Spin-split bands, Valley, Strain~~~~~~~~~~~~~~~~~~~~~~Email:sandongyuwang@163.com}

\maketitle

\section{Introduction}
The  antiferromagnetic (AFM)  materials do not have any net magnetic moment, which are robust to external
magnetic perturbation,  and  have  ultra-high dynamic speed,  allowing
high-speed device operation\cite{k1,k2}. However,  it is often hard to realize spin-polarized currents in collinear AFM
systems because of the missing spin-splitting  in the band
structures\cite{k3}. Recently, the spin-splitting has been achieved  in collinear symmetry-compensated antiferromagnets, be called
altermagnetism\cite{k4,k5,k6}.  The spin-splitting is only originated from the simple AFM
order with special magnetic space group, and  the relativistic spin-orbital coupling (SOC)  is thus not needed.
Several bulk materials have been predicted to be  altermagnetism, such as $\mathrm{RuO_2}$\cite{k7}, $\mathrm{FeF_2}$\cite{k7-1}, MnTe\cite{k7-2},
some organic AFMs\cite{k8}, $\mathrm{MnF_2}$\cite{k9} and some $\mathrm{GdFeO_3}$-type
perovskites\cite{k10}.

Besides, several two-dimensional (2D)  materials
have also been predicted to have this spin-splitting with special AFM order, such as $\mathrm{Cr_2O_2}$\cite{k11,k12} and  $\mathrm{V_2Se_2O}$\cite{k13}. However,
these 2D  altermagnetic materials have inversion symmetry, leading to missing  piezoelectricity. Searching for  2D piezoelectric altermagnetism (PAM) may be significative and challenging. In fact, many 2D  piezoelectric ferromagnetisms (PFMs) has been predicted\cite{q15-0,q15-1,q15-2,q15-3,q15-4}, which achieve the combination of piezoelectricity and ferromagnetic (FM) order in a 2D material.
Most of these 2D PFMs possess Janus structure, which can destroy  out-of-plane symmetry, inducing  out-of-plane piezoelectricity. Therefore, 2D Janus materials may provide a potential platform for  searching 2D PAMs. Another question is how to achieve spin polarization in 2D altermagnetic materials, and strain may be a very effective way to induce spin polarization\cite{k11,k12,k13}.

Here,  we construct  Janus monolayer  $\mathrm{Cr_2SO}$ as derivative of altermagnetic $\mathrm{Cr_2O_2}$ monolayer. Calculated results show that $\mathrm{Cr_2SO}$ is still a 2D altermagnetic material, and has large out-of-plane piezoelectricity, which achieves PAM. Both spin and valley polarizations can be realized by simply breaking the corresponding crystal symmetry with uniaxial strain. Compared with existing 2D  altermagnetic  $\mathrm{Cr_2O_2}$\cite{k11,k12} and  $\mathrm{V_2Se_2O}$\cite{k13},  $\mathrm{Cr_2SO}$ has two main differences: (1) it has large out-of-plane piezoelectricity; (2) near the Fermi level, the states of conduction and valence bands  around X or Y high-symmetry point are  with opposite spins.

\section{Computational detail}
 We perform the spin-polarized  first-principles calculations within density functional theory (DFT)\cite{1}by using the standard VASP package\cite{pv1,pv2,pv3} within the projector augmented-wave (PAW) method. The generalized gradient
approximation  of Perdew-Burke-Ernzerhof (PBE-GGA)\cite{pbe} is used as the exchange-correlation functional.
 We set kinetic energy cutoff  of 500 eV,  total energy  convergence criterion of  $10^{-8}$ eV, and  force convergence criterion of 0.0001 $\mathrm{eV.{\AA}^{-1}}$.
To account for electron correlation of Cr-3$d$ orbitals, a Hubbard correction $U_{eff}$=3.55 eV\cite{k11,k12}  is employed within the
rotationally invariant approach proposed by Dudarev et al.
The out-of-plane interaction is avoided by taking a vacuum of more than 16 $\mathrm{{\AA}}$.
  We use a 18$\times$18$\times$1 Monkhorst-Pack k-point meshes to sample the Brillouin zone (BZ) for calculating electronic structures, elastic  and piezoelectric properties.

The elastic stiffness tensor  $C_{ij}$   are calculated by using strain-stress relationship (SSR) method. The piezoelectric stress tensor $e_{ij}$  are calculated  by density functional perturbation theory (DFPT) method\cite{pv6}. The  $C^{2D}_{ij}$/$e^{2D}_{ij}$ have been renormalized by   $C^{2D}_{ij}$=$L_z$$C^{3D}_{ij}$/$e^{2D}_{ij}$=$L_z$$e^{3D}_{ij}$, where the $L_z$ is  the length of unit cell along $z$ direction.
Based on  finite displacement method, the interatomic force constants (IFCs)  are calculated by using  5$\times$5$\times$1 supercell, and the phonon dispersion spectrum  is obtained by the  Phonopy code\cite{pv5}.
The  ab-initio molecular dynamics (AIMD) simulations  using NVT ensemble are performed   for more than
8000 fs with a time step of 1 fs by using a 4$\times$4$\times$1 supercell. The elastic, piezoelectric, phonon and AIMD calculations are all carried out with altermagnetic order.

\begin{figure}
  \includegraphics[width=8cm]{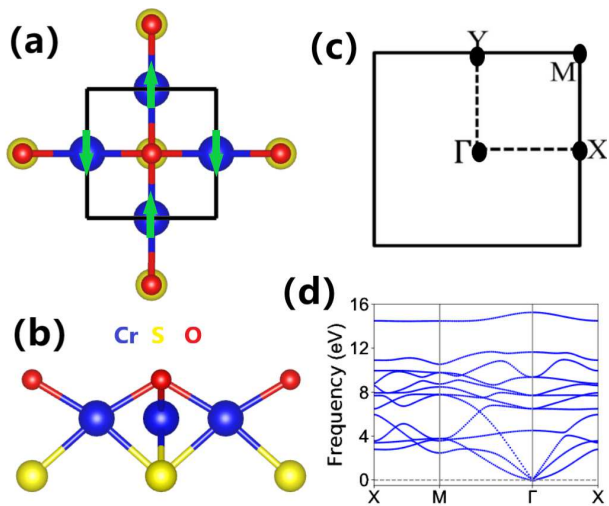}
  \caption{(Color online) For Janus monolayer $\mathrm{Cr_2SO}$, (a) and (b):  top and side views of the  crystal structures; (c):  The first BZ with high symmetry points; (d):the phonon dispersion curves.}\label{st}
\end{figure}

\section{Crystal structure and stability}
Monolayer $\mathrm{Cr_2SO}$ and $\mathrm{Cr_2O_2}$ have similar crystal structures, and they contain three atomic sublayers with two co-planar Cr atoms as  middle layer and  O/S atoms as upper and lower layers.
The Janus monolayer   $\mathrm{Cr_2SO}$ can be constructed  by  replacing one of two O  layers with S atoms in monolayer  $\mathrm{Cr_2O_2}$.
 The schematic crystal structures of $\mathrm{Cr_2SO}$ are shown in \autoref{st} (a) and (b).  The  monolayer  $\mathrm{Cr_2O_2}$  possesses  $P4/mmm$ space group (No.123)\cite{k11,k12}, and
 the monolayer  $\mathrm{Cr_2SO}$  crystallizes in the $P4mm$ space group (No.99). For monolayer $\mathrm{Cr_2O_2}$,
  the key space-group symmetry operations  contain
space inversion $P$, $C_4$ rotation and $M_x$/$M_y$/$M_z$/$M_{xy}$ mirror, indicating no piezoelectricity. With respect to $\mathrm{Cr_2O_2}$, the  $P$ and $M_z$ of $\mathrm{Cr_2SO}$  are removed, which means that the $\mathrm{Cr_2SO}$ will be piezoelectric.
In one unit cell, the FM and AFM configurations are constructed to determine magnetic ground state and lattice constants.
Calculated results show that the AFM  configuration is ground state of $\mathrm{Cr_2SO}$, and its energy is 1.122 eV lower than that of FM case within GGA+$U$. The optimized lattice constants $a$=$b$=3.66 $\mathrm{{\AA}}$ by GGA+$U$ for AFM case.
The magnetic easy-axis is investigated by calculating the energy
difference of the magnetization orientation along the (100)
and (001) cases, which is defined as magnetic anisotropy energy (MAE).
Calculated results show that the MAE is -93 $\mathrm{\mu eV}$/Fe, which indicates that the easy-axis of $\mathrm{Cr_2SO}$ is in-plane.

To evaluate the stability of $\mathrm{Cr_2SO}$,
the phonon dispersion, molecular dynamics and elastic
constants are calculated by using GGA+$U$ for AFM case. The phonon
dispersions of $\mathrm{Cr_2SO}$ are shown in \autoref{st} (d), and no imaginary frequencies can be observed,  indicating its dynamic stability.
To corroborate the thermal stability, the evolution of total
energy vs time are calculated using AIMD at 200 and 300 K, which are  shown in FIG.1  and FIG.2 of electronic supplementary information (ESI).
At 200 K, the energies are kept stable, and the  crystal
features are preserved after 8 ps,  which confirms its thermal stability.  However, at 300 K, the thermal stability of  $\mathrm{Cr_2SO}$ is broken.
The independent elastic constants $C_{11}$, $C_{12}$ and  $C_{66}$ of  $\mathrm{Cr_2SO}$ are 64.93 $\mathrm{Nm^{-1}}$, 37.26 $\mathrm{Nm^{-1}}$ and 30.05 $\mathrm{Nm^{-1}}$,  which  satisfy the  Born  criteria of mechanical stability:
$C_{11}>0$, $C_{66}>0$, $C_{11}-C_{12}>0$,  confirming  its mechanical stability.

\section{electronic structures}
\begin{figure}
  \includegraphics[width=8cm]{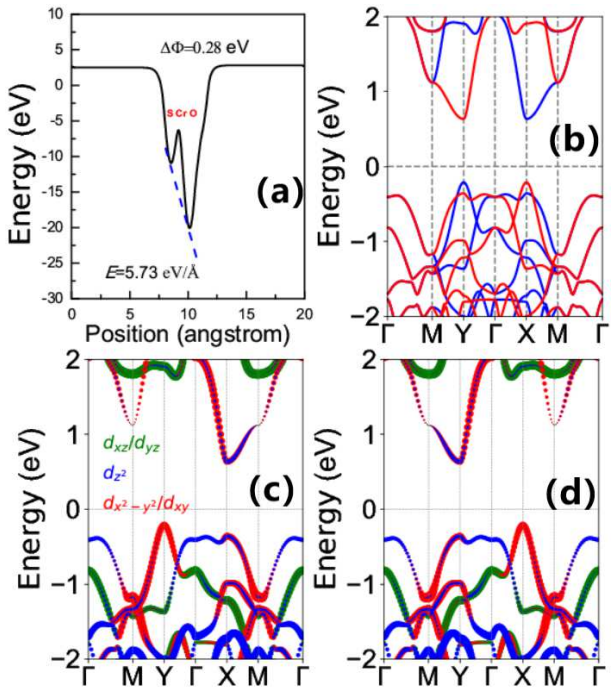}
\caption{(Color online)For $\mathrm{Cr_2SO}$, (a):planar averaged electrostatic potential energy variation along $z$ direction. $\Delta\varphi$ is the potential energy difference across the layer. $E$ stands for the intrinsic polar field; (b): the energy
band structures. The spin-up
and spin-down channels are depicted in blue and red; (c) and (d):the Cr-$d$-resolved band structures for spin-up
and spin-down channels. }\label{t1}
\end{figure}

The planar average of the electrostatic potential energy along out-of-plane direction is
shown in \autoref{t1} (a). Due to  mirror asymmetry, an electrostatic potential gradient ($\Delta \Phi$) of about 0.28 eV
 is produced, which is
related to the work function change of the structure.
 Due to the
electron redistribution, there is an inherent electric field with the magnitude
of about  5.73 $\mathrm{V/{\AA}}$,  implying a very
strong vertical polarization in the $\mathrm{Cr_2SO}$ monolayer.

The Cr atoms in $\mathrm{Cr_2SO}$ monolayer possess a local magnetic moment around 3.46 $\mu_B$, and
the two Cr-atom sublattices with   the 2D $\mathrm{N\acute{e}el}$ AFM order are related by $M_{xy}$ mirror symmetry  but
cannot be transformed to each other by any translation operation, which leads to existing altermagnetism. The energy band structures of $\mathrm{Cr_2SO}$ are plotted in \autoref{t1} (b) by GGA+U method without SOC. There are two valleys at X and Y high-symmetry points for both conduction and valence bands, which are  related by the mirror symmetry. States around X and Y points are mainly from two different Cr atoms with opposite spins due to the $\mathrm{N\acute{e}el}$ AFM order, producing altermagnetism in the absence of SOC and  spin-valley locking. Compared with spin-gapless $\mathrm{Cr_2O_2}$\cite{k11,k12},  it is clearly seen that   $\mathrm{Cr_2SO}$ is  a direct band gap semiconductor with gap value of 0.838 eV, which may be due to strong inherent electric field ($\thicksim$5.73 $\mathrm{V/{\AA}}$).   The valence band maximum (VBM) and conduction band bottom (CBM) of $\mathrm{Cr_2SO}$ are at high symmetry X or Y point, while those of $\mathrm{Cr_2O_2}$ deviate from X or Y point.
Near the Fermi level, states of conduction and valence bands  around X or Y point are from two different Cr atoms with opposite spins,  while those of $\mathrm{Cr_2O_2}$ are  with the same spins\cite{k11,k12}. For $\mathrm{V_2Se_2O}$ monolayer, the same spins are also observed for  states of conduction and valence bands  around X or Y point near the Fermi level\cite{k13}.
 According to \autoref{t1} (b) and (d), it is found that $d_{x^2-y^2}$+$d_{xy}$+$d_{z^2}$ orbitals  dominate X and Y valleys of conduction bands, and the X and Y valleys of valence bands are mainly from  $d_{x^2-y^2}$+$d_{xy}$ orbitals.

The electronic correlation effects on electronic structures of $\mathrm{Cr_2SO}$ are considered by using different $U$ values.
Firstly, the lattice constants $a$  at different $U$ (0-4 eV) are optimized, and then calculate electronic properties of $\mathrm{Cr_2SO}$. With increasing $U$, the lattice constants $a$ increases, and  its change  is  about 0.232 $\mathrm{{\AA}}$. According to FIG.3 of ESI,
it is found that $\mathrm{Cr_2SO}$ is always a AFM ground state.
The evolutions of energy  band
structures as a function of $U$ are calculated, which are plotted in FIG.4 of ESI.
With increasing the $U$, monolayer $\mathrm{Cr_2SO}$ experiences a phase transition from metal to semiconductor, and the critical point is about 2 eV. Similar phenomenon can be also observed in monolayer $\mathrm{V_2Se_2O}$\cite{k13}.
For $U$$\geqslant$2 eV, near the Fermi level, states of conduction and valence bands  around X or Y point are always from two different Cr atoms with opposite spins.

\begin{figure*}
  \includegraphics[width=16cm]{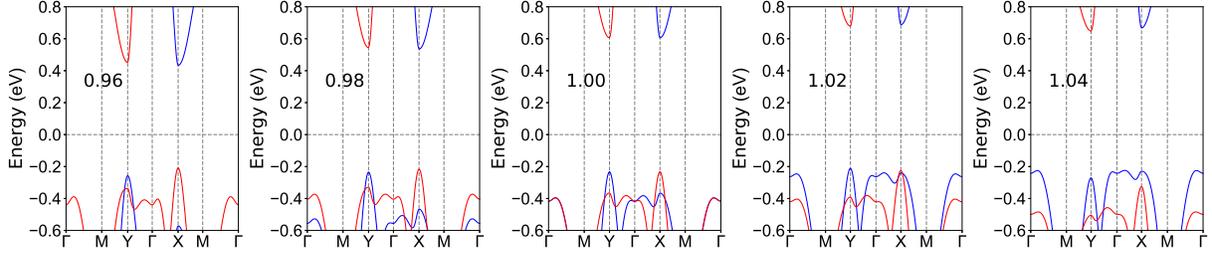}
\caption{(Color online) The energy  band structures of  $\mathrm{Cr_2SO}$  at representative $a/a_0$. The compressive strain makes $X$ and $Y$ valleys  of both conduction and valence bands separate from other bands.}\label{band}
\end{figure*}

\begin{figure}
  \includegraphics[width=8cm]{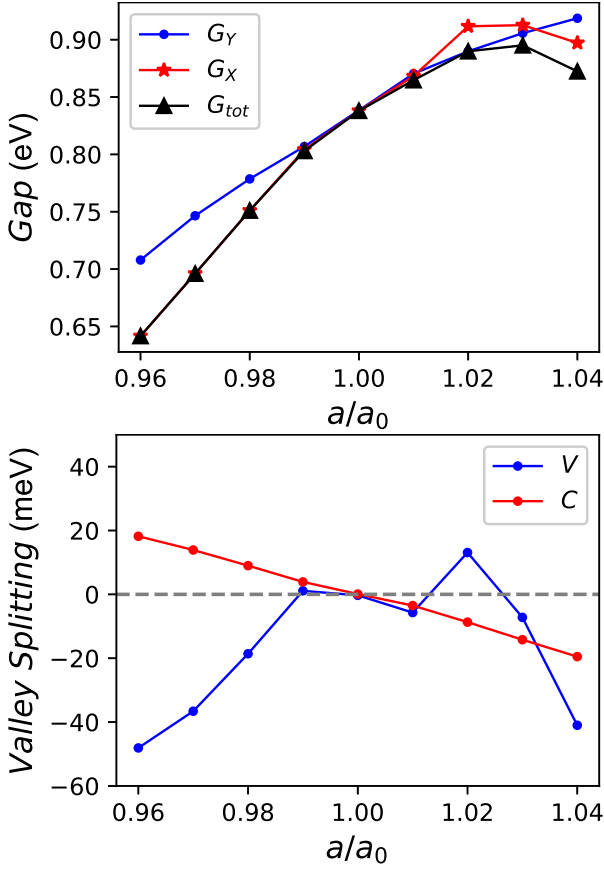}
  \caption{(Color online)For $\mathrm{Cr_2SO}$, the related  band gaps  including  the  global gap [$G_{tot}$] and gaps of Y and X valleys [$G_{Y}$ and $G_{X}$] (top panel),  and   valley splitting  for both valence [$V$] and condition [$C$] bands (bottom panel) as a function of $a/a_0$. }\label{gap}
\end{figure}

\section{Piezoelectric properties}
Due to  inversion symmetry, the $\mathrm{Cr_2O_2}$ monolayer possesses no piezoelectricity. However,
the monolayer $\mathrm{Cr_2SO}$  has piezoelectricity with  missing  $P$ and $M_z$ symmetries.
Namely,  the reflectional
symmetry of $\mathrm{Cr_2SO}$  disappears across the $xy$ plane, but holds across the $xz$ or $yz$ plane, which leads to  existing out-of-plane piezoelectricity and missing in-plane piezoelectricity.  The third-rank piezoelectric stress tensor  $e_{ijk}$ and strain tensor $d_{ijk}$ can be used to describe the piezoelectric response of a material.  They can be expressed as the sum of ionic and electronic contributions:
 \begin{equation}\label{pe0}
 \begin{split}
      e_{ijk}=\frac{\partial P_i}{\partial \varepsilon_{jk}}=e_{ijk}^{elc}+e_{ijk}^{ion}\\
   d_{ijk}=\frac{\partial P_i}{\partial \sigma_{jk}}=d_{ijk}^{elc}+d_{ijk}^{ion}
   \end{split}
 \end{equation}
Where $P_i$, $\varepsilon_{jk}$ and $\sigma_{jk}$ are polarization vector, strain and stress, respectively.  The superscripts $elc$ and $ion$  mean electronic and ionic contributions. The  $e_{ijk}^{elc}$ and $d_{ijk}^{elc}$ ($e_{ijk}$ and $d_{ijk}$)   are also called clamped-ion (relaxed-ion)  piezoelectric coefficients. The $e_{ijk}$ is related with  $d_{ijk}$  by elastic tensor $C_{mnjk}$:
 \begin{equation}\label{h1}
    e_{ijk}=\frac{\partial P_i}{\partial \varepsilon_{jk}}=\frac{\partial P_i}{\partial \sigma_{mn}}.\frac{\partial \sigma_{mn}}{\partial\varepsilon_{jk}}=d_{imn}C_{mnjk}
 \end{equation}

By using  Voigt notation,  only considering the in-plane strain and stress\cite{o2}, the \autoref{h1} can be reduced into:
  \begin{equation}\label{h2}
 \left(
    \begin{array}{ccc}
    0 & 0 & 0 \\
     0 & 0 & 0 \\
      e_{31} & e_{31} & 0 \\
    \end{array}
  \right)
  =\left(
    \begin{array}{ccc}
       0 &  0 & 0 \\
       0 & 0 &  0 \\
      d_{31} & d_{31} &0 \\
    \end{array}
  \right)
  \left(
    \begin{array}{ccc}
      C_{11} & C_{12} & 0 \\
     C_{12} & C_{11} &0 \\
      0 & 0 & C_{66} \\
    \end{array}
  \right)
    \end{equation}
The existing  $e_{31}$/$d_{31}$  means that only vertical piezoelectric polarization can be induced, when a  uniaxial  strain is applied. By solving the \autoref{h2}, the $d_{31}$  can be obtained:
\begin{equation}\label{pe2-2}
    d_{31}=\frac{e_{31}}{C_{11}+C_{12}}
\end{equation}

The primitive cell is used   to calculate the  $e_{31}$  of  monolayer $\mathrm{Cr_2SO}$. The calculated $e_{31}$ is -0.987$\times$$10^{-10}$  with ionic part 0.111$\times$$10^{-10}$ and electronic part -1.098$\times$$10^{-10}$. The electronic and ionic contributions  have  opposite signs, and   the electronic part dominates the  piezoelectricity.  And then, the  $d_{31}$  can be attained from \autoref{pe2-2}, and the corresponding value is -0.97 pm/V.  The minus sign depends on the choice of coordinate system. If the  $\pm z$ directions are reversed, the $d_{31}$ will become 0.97 pm/V.
A large out-of-plane piezoelectric response ($d_{31}$) is very important
to be compatible with the
nowadays bottom/top gate technologies.  The $|d_{31}|$  of  $\mathrm{Cr_2SO}$ is compared with or  higher  than ones of many known 2D  materials, like
functionalized h-BN (0.13 pm/V)\cite{o1}, kalium decorated graphene (0.3
pm/V)\cite{o2},  the oxygen functionalized MXenes (0.40-0.78 pm/V)\cite{q9},  Janus group-III materials (0.46 pm/V)\cite{q7-6}, Janus BiTeI/SbTeI  monolayer (0.37-0.66 pm/V)\cite{o3}, Janus monolayer transition metal dichalcogenides (0.03 pm/V)\cite{o3-1} and $\alpha$-$\mathrm{In_2Se_3}$
(0.415 pm/V)\cite{o4}. The large $|d_{31}|$ may be  related to large electronegativity difference of S and O atoms\cite{q15-3,q15-4}.
The large $|d_{31}|$  provide possibility to tune spin-split bands of   altermagnetic  $\mathrm{Cr_2SO}$ by piezoelectric effect.

\section{uniaxial strain induces spin-valley polarization}
To induce spin-valley polarization in $\mathrm{Cr_2SO}$, an experimentally feasible approach is to destroy $M_{xy}$ symmetry by  applying  uniaxial strain along $x$ or $y$ direction, which will lead to nonequivalent X and Y valleys, giving rise to spin-valley polarization.  We use $a/a_0$ (0.96 to 1.04)  to simulate the uniaxial strain along $x$  direction, and  the lattice constants $b$ along $y$ direction is optimized.
 The  $a$ and $a_0$ are the strained and  unstrained lattice constants with $a/a_0$$<$1 ($a/a_0$$>$1) being compressive (tensile) case. If the strain is applied along $y$ direction, the opposite gap change and spin-valley
polarization are exactly generated,  since two valleys are related with $M_{xy}$ symmetry.
The Young's modulus is calculated to elucidate
mechanical performance of $\mathrm{Cr_2SO}$, which is plotted in FIG.5 of ESI. The calculated  $C_{2D}$ of  $\mathrm{Cr_2SO}$ along $x$ direction (44 $\mathrm{Nm^{-1}}$) is very small than those of many known 2D materials, such as graphene ($\sim 340\pm 40$ Nm$^{-1}$) and MoS$_2$ ($\sim 126.2$ Nm$^{-1}$)~\cite{q5-1,q5-1-1}, indicating its better mechanical flexibility.

 The energy differences ($\Delta E$) between AFM  and  FM states  vs $a/a_0$ are calculated to determine
 the magnetic ground state of strained  $\mathrm{Cr_2SO}$, which are plotted in FIG.6 of ESI.
Within considered strain range,  the  $\Delta E$ is always negative,  confirming that strained  $\mathrm{Cr_2SO}$ is  AFM ground state.
The energy band structures of strained monolayer $\mathrm{Cr_2SO}$  are shown in \autoref{band}.
The evolutions of related energy band gap (global gap [$G_{tot}$] and gaps of Y and X valleys [$G_{Y}$ and $G_{X}$]) and the valley splitting ($\Delta E_C=E_{Y}^C-E_{X}^C$ and $\Delta E_V=E_{Y}^V-E_{X}^V$) for both valence and condition bands  as a function of $a/a_0$ are plotted in \autoref{gap}.
For unstrained  case, the  global gap  and gaps of $X$ and $Y$ valleys are the same, indicating no valley polarization.
It is found that both compressive and tensile strains can induce unequal gap between  $X$ and $Y$ valleys, which can produce valley polarization for both conduction and valence bands. It is clearly seen that compressive strain is in favour of separating $X$ and $Y$ valleys  of both conduction and valence bands from other bands, which makes for  easily  manipulating these valleys in experiment.
 At $a/a_0$=0.96, the corresponding valley splitting are 18 meV  (-48 meV) for conduction (valence) band.

\begin{figure}
  \includegraphics[width=8cm]{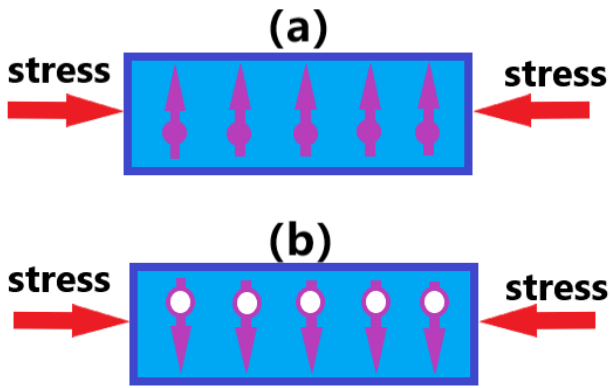}
  \caption{(Color online)For $\mathrm{Cr_2SO}$ with compressive strain, an appropriate electron/hole doping
 leads to that only one valley has doped electron/holes.
 Due to spin-valley locking, the electron carriers possess spin-up character (a), while hole carriers have spin-down one (b).  }\label{qj}
\end{figure}

For  compressive strain, an appropriate electron/hole doping
can move the Fermi level to fall between the X and Y
valleys, which leads to that only one valley has doped electron/holes.
 Due to spin-valley locking (see \autoref{band} and \autoref{qj}), the electron carriers possess spin-up character, while hole carriers have spin-down one.   Upon appropriate doping,  the magnetization can be produced due to the polarized carriers. The electron and hole doping leads to opposite magnetization direction.  For a given electron/hole density, the  magnetization can be tuned by uniaxial strain.
These provide a platform for spin device applications.

\section{Conclusion}
In summary,  we propose a 2D Janus  altermagnetic material $\mathrm{Cr_2SO}$, which is dynamically, mechanically and thermally  stable.
Calculated results show that $\mathrm{Cr_2SO}$ is a semiconductor with spin-split bands near the Fermi level.
The structural symmetry-breaking and large electronegativity difference of O and S atoms lead to a large
out-of-plane piezoelectric coefficient ($|d_{31}|$) of 0.97 pm/V.
The $\mathrm{Cr_2SO}$ possesses spin-valley locking, which is comprised of spin-polarized
valleys related by a crystal symmetry. The spin-valley polarization can be induced by simply breaking the corresponding crystal
symmetry with uniaxial strain. In fact, we provide a very effective method to induce piezoelectricity from centrosymmetric altermagnetism by constructing Janus structure. The method, analysis and results  can be readily extended to other  members of centrosymmetric monolayer altermagnetism  AB (A = Co, Cr, Fe, Mn, Mo, Nb, Ni, Pd, Rh, Ru, Sc, Ti, V, Y,
Zr; B = B, C, N, O, F, Al, Si, P, S, Cl, Ga, Ge, As, Se, Br), possessing the same structure and AFM magnetic configuration with $\mathrm{Cr_2O_2}$\cite{k11}. Based on these altermagnetic monolayers, constructing Janus structure can  realize piezoelectric altermagnetism.

\begin{acknowledgments}
This work is supported by Natural Science Basis Research Plan in Shaanxi Province of China  (2021JM-456). We are grateful to Shanxi Supercomputing Center of China, and the calculations were performed on TianHe-2..
\end{acknowledgments}

\end{document}